\documentclass[11pt,twoside]{article}
\usepackage[hang,flushmargin]{footmisc} 
\usepackage{subfigure}
\usepackage{graphicx}
\usepackage{amsmath}
\usepackage{color}
\usepackage{listings}
\usepackage{xspace}

\definecolor{blue}{rgb}{0.098,0.357,0.675}
\definecolor{green}{rgb}{0.5,0.75,0.0}
\definecolor{gray}{rgb}{0.25,0.25,0.25}
\definecolor{ngreen}{rgb}{0.42, 0.647, 0.004} 

\lstdefinelanguage{CUDA}[]{C++}{
    morekeywords={__global__,__device__,__shared__,__syncthreads,threadIdx,blockIdx,float3,float4,rsqrtf},
}
\usepackage[
    colorlinks, linkcolor={blue},citecolor={blue},urlcolor={blue}
    ]{hyperref}

\makeatletter
\def\url@leostyle{%
  \@ifundefined{selectfont}{\def\UrlFont{\sf}}{\def\UrlFont{\small\ttfamily}}}
\makeatother
\newcommand{\bigON}{$\mathcal{O}(N)$}
\newcommand{\bigONlogN}{$\mathcal{O}(N\log N)$}
\newcommand{\bigOsq}{$\mathcal{O}(N^2)$}
\newcommand{\fmm}{\textsc{FMM}\xspace}
\newcommand{\bem}{\textsc{BEM}\xspace}

\newcommand{\cpu}{\textsc{CPU}}
\newcommand{\gpu}{\textsc{GPU}}
\newcommand{\cuda}{\textsc{CUDA}\xspace}
\newcommand{\sdk}{\textsc{SDK}}
\newcommand{\sse}{\textsc{SSE}}
\newcommand{\mac}{\textsc{MAC}}

\newcommand{\MM}{\textsc{M}2\textsc{M}}
\newcommand{\PM}{\textsc{P}2\textsc{M}}
\newcommand{\ML}{\textsc{M}2\textsc{L}}
\newcommand{\LL}{\textsc{L}2\textsc{L}}
\newcommand{\LP}{\textsc{L}2\textsc{P}}
\newcommand{\MP}{\textsc{M}2\textsc{P}}
\newcommand{\PP}{\textsc{P}2\textsc{P}}
\newcommand{\grape}{\textsc{GRAPE}\xspace}

\newlength{\up}
\newlength{\hup}
\title{\Huge\raggedright\sffamily\bfseries\color{blue}Treecode and fast multipole method for N-body simulation with CUDA}
\date{}
\author{}

\begin{document}
\maketitle
\vspace{-5mm}

{\large\bfseries\itshape Rio Yokota}\\
{\itshape \small Boston University}
\bigskip

{\large\bfseries\itshape Lorena A. Barba}\\
{\itshape \small Boston University}

\section{Introduction}
\vspace{\hup}

The classic $N$-body problem refers to determining the motion of $N$ particles that interact via a long-distance force, such as gravitation or electrostatics.  A straightforward approach to obtaining the forces affecting each particle is the evaluation of all pair-wise interactions, resulting in \bigOsq\ computational complexity.  This method is only reasonable for moderate-size systems, or to compute near-field interactions, in combination with a far-field approximation.  In the previous \emph{GPU Gems} volume \cite{gems3}, the acceleration of the all-pairs computation on \gpu s was presented for the case of the gravitational potential of $N$ masses.  The natural parallelism available in the all-pairs kernel allowed excellent performance on the \gpu\ architecture, and the direct kernel of \cite{NylandHarrisPrins2007} achieved over 200 Gigaflops on the \textsc{G}e\textsc{F}orce \textsc{8800 GTX}, calculating more than 19 billion interactions per second with $N=$16,384.  In the present contribution, we have addressed the more involved task of implementing the fast $N$-body algorithms that are used for providing a far-field approximation:  the \bigONlogN\ treecode \cite{BarnesHut1986} and \bigON\ fast multipole method \cite{GreengardRokhlin1987}.

Before embarking on the presentation of the algorithms and how they are efficiently cast onto the \gpu, let us give some context.  The $N$-body problem of astrophysics was such a strong motivator to computational science, that it drove creation of a special supercomputer in Japan.  The history of this massively successful series of machines, called \grape, is summarized in the book by its creators, \cite{MakinoTaiji1998}; a popular science magazine article also gives an overview \cite{Taubes1997}. The \grape machines continued to break records into the $21^{st}$ century, but the size of the problems they can tackle using the \bigOsq\  all-pairs force evaluation is still limited by the computational complexity.  As stated in \cite{BoardSchulten2000}, \emph{``complexity trumps hardware.''}

Clever algorithms can have a drastic impact on the capabilities of computational science to solve challenging problems.  A case in point is the fast Fourier transform, which has enabled a variety of successful research areas (\emph{e.g.}, the triumph of spectral methods in the simulation of turbulence).  The first viable fast algorithms for $N$-body problems \cite{Appel1985,BarnesHut1986} combined two ideas:  \emph{(i)}  approximating the effect of a group of distant particles (charges, or masses) by their first few moments, and \emph{(ii)} rationally dividing space in a hierarchical fashion to establish acceptable margins of distance for these approximations. These two ideas combined in an algorithm result in the so-called \emph{treecode}, reducing the computational complexity to \bigONlogN. The critical third idea that was introduced in the fast multipole method, \fmm, is the ``local expansion''.  This mathematical representation allows groups of distant particles to interact with \emph{groups} of targets, thereby reducing the complexity further to the ideal \bigON\ scaling. One essential difference between treecodes and \fmm that remains is the method of achieving a desired accuracy in the approximations.  Treecodes ensure a given accuracy by restricting the acceptable distances for group-to-target interactions, while \fmm looks to the series representation and chooses a proper truncation for specified accuracy.

The advantage of fast algorithms was appreciated by the \grape team early on; a modified treecode by \cite{Barnes1990} was first used in combination with the \grape hardware by \cite{Makino1991}, and continued in later generations of the machine \cite{MakinoTaiji1995,KawaiETal1999}. The hardware architecture limited the order of the multipole expansions to only the dipole term, however, which motivated the development of a new algorithm: the pseudo-particle method \cite{Makino99}.  Thus, the interesting history of the \grape project illustrates well the interplay between architecture and algorithms.  In fact, there are many parallels with \gpu s, as used for general-purpose scientific computing. We are reminded here of the statement in \cite{TrefethenBau1997}:  

\begin{quotation}
\noindent ``the fundamental law of computer science [is]: the faster the computer, the greater the importance of speed of algorithms.''
\end{quotation}

Fast algorithms for $N$-body problems have diverse practical applications. We have mentioned astrophysics, the paradigm problem. Of great importance is also the calculation of electrostatic (Coulomb) interactions of many charged ions in biological molecules. Proteins and their interactions with other molecules constitute a great challenge for computation, and fast algorithms can enable studies at physiologically relevant scales \cite{BoardETal1992,SaguiDarden1999}.  Both gravitational and electrostatic problems are mathematically equivalent to solving a Poisson equation for the scalar potential.  A general method for the solution of Poisson problems in integral form is described in \cite{GreengardLee1996}, using the \fmm in a very interesting way to patch local solutions.  In \cite{EthridgeGreengard2001}, instead, the \fmm is applied directly to the volume integral representation of the Poisson problem.  These general Poisson solvers based on \fmm open the door to using the algorithm in various situations where complex geometries are involved, such as fluid dynamics, and also shape representation and recognition \cite{GorelickETal2006}.  

The  \fmm for the solution of Helmholtz equations was first developed in \cite{Rokhlin1990}, and is explained in great detail in the book by \cite{GumerovDuraiswami2004}. The integral-equation formulation is an essential tool in this context, reducing the volumetric problem into one of an integral over a surface.  The \fmm allows fast solution of these problems by accelerating the computation of dense matrix-vector products arising from the discretization of the integral problem.   In fact, the capability of boundary element methods, \bem,  is in this way significantly enhanced; see \cite{Nishimura2002} and \cite{LiuNishimura2006a}.  These developments make possible the use of the \fmm for many physical and engineering problems, such as seismic, magnetic and acoustic scattering \cite[\emph{e.g.},][]{Fujiwara1998,DonepudiETal2003,DarveHave2004b,GumerovDuraiswami2009}.   The recent book by \cite{Liu2009} covers applications in elastostatics, Stokes flow, and acoustics;  some notable applications including acoustic fields of building and sound barrier combinations, and also a wind turbine model, were presented in \cite{BapatETal2009}.

Due to the variety and importance of applications of treecodes and \fmm, the combination of algorithmic acceleration with hardware acceleration can have tremendous impact.  Alas, programming these algorithms efficiently is no piece of cake.  
In this contribution, we aim to present \gpu\ kernels for treecode and \fmm in, as much as possible, an uncomplicated, accessible way.  The interested reader should consult some of the copious literature on the subject for a deeper understanding of the algorithms themselves.  Here, we will offer the briefest of summaries.  We will focus our attention on achieving a \gpu\ implementation that is efficient in its utilization of the architecture, but without applying the most advanced techniques known in the field (which would complicate the presentation). These advanced techniques that we deliberately did not discuss in the present contribution are briefly summarized in section \ref{advanced}, for completeness.  Our target audience is the researcher involved in computational science with an interest in using fast algorithms for any of the applications mentioned above:  astrophysics, molecular dynamics, particle simulation with non-negligible far fields, acoustics, electromagnetics, and boundary integral formulations.


\vspace{\up}
\section{Fast N-body simulation} 
\vspace{\hup}

As in  \cite{NylandHarrisPrins2007}, we will use as our model problem the calculation of the gravitational potential of $N$ masses. We have the following expressions for the potential and force, respectively,  on a body $i$:

\vspace{-1.5em}

\begin{equation}
\Phi_i=m_i\sum_{j=1}^{N} \frac{m_j}{r_{ij}}, \qquad
\mathbf{F}_i=-\nabla\Phi_i
\label{eq:force}
\end{equation}

Here, $m_i$ and $m_j$ are the masses of bodies $i$ and $j$, respectively; and $\mathbf{r}_{ij} = \mathbf{x}_j -\mathbf{x}_i$ is the vector 
from body $i$ to body $j$. Since the distance vector $\mathbf{r}_{ij}$ is a function of both $i$ and $j$, an all-pairs summation must be performed. This results in \bigOsq\ computational complexity. In the treecode, the sum for the potential is factored into a near-field and a far-field expansion, in the following way,

\begin{equation}
\Phi_i=\sum_{n=0}^{\infty}\sum_{m=-n}^{n}m_ir_{i}^{-n-1}Y_{n}^{m}(\theta_i,\phi_i)\sum_{j=1}^{N}\underbrace{m_j\rho_{j}^{n}Y_{n}^{-m}(\alpha_j,\beta_j)}_{M_{n}^{m}}.
\label{eq:multipole}
\end{equation}

Calculating the summation for $M_{n}^{m}$ in this manner can be interpreted as the clustering of particles in the far field. In the above expression, $Y_{n}^{m}$ is the spherical harmonic function, and $(r,\theta,\phi)$; $(\rho,\alpha,\beta)$ are the distance vectors from the center of the expansion to bodies $i$ and $j$, respectively. The key is to factor the all-pairs interaction into a part that involves only $i$, and a part that involves only $j$, hence allowing the summation of $j$ to be performed outside of the loop for $i$.  The condition $\frac{\rho}{r}<1$, which is required for the series expansion to converge, prohibits the clustering of particles in the near field. Therefore, a tree structure is used to form a hierarchical list of $\log N$ cells that interact with $N$ particles. This results in \bigONlogN\ computational complexity. 

The complexity can be further reduced by considering cluster-to-cluster interactions \footnote{ The groups or clusters of bodies reside in a sub-division of space for which various authors use the term ``box'' or ``cell''; \emph{e.g.}, ``leaf-cell'' as used in \cite{NylandHarrisPrins2007} corresponds to the smallest sub-domain.}. In the \fmm, a second series expansion is used for such interactions:

\begin{equation}
\Phi_i=\sum_{n=0}^{\infty}\sum_{m=-n}^{n}m_ir_{i}^{n}Y_{n}^{m}(\theta_i,\phi_i)\sum_{j=1}^{N}\underbrace{m_j\rho_{j}^{-n-1}Y_{n}^{-m}(\alpha_j,\beta_j)}_{L_{n}^{m}},
\label{eq:local}
\end{equation}
\noindent where the near-field expansion and far-field expansion are reversed. The condition for this expansion to converge is  $\frac{r}{\rho}<1$, which means that the clustering of particles using $L_{n}^{m}$ is only valid in the near field. The key here is to translate multipole expansion coefficients $M_{n}^{m}$ of cells in the far field  to local expansion coefficients $L_{n}^{m}$ of cells in the near field, resulting in a cell-cell interaction. Due to the hierarchical nature of the tree structure, each cell needs to only consider the interaction with a constant number of neighboring cells. Since the number of cells is of \bigON, the \fmm has a complexity of \bigON. Also, it is easy to see that keeping the number of cells proportional to $N$ results in an asymptotically constant number of particles per cell. This prevents the direct calculation of the near field from adversely affecting the asymptotic behavior of the algorithm.

\begin{figure}[t]
\begin{center}
\includegraphics[width=1.0\textwidth]{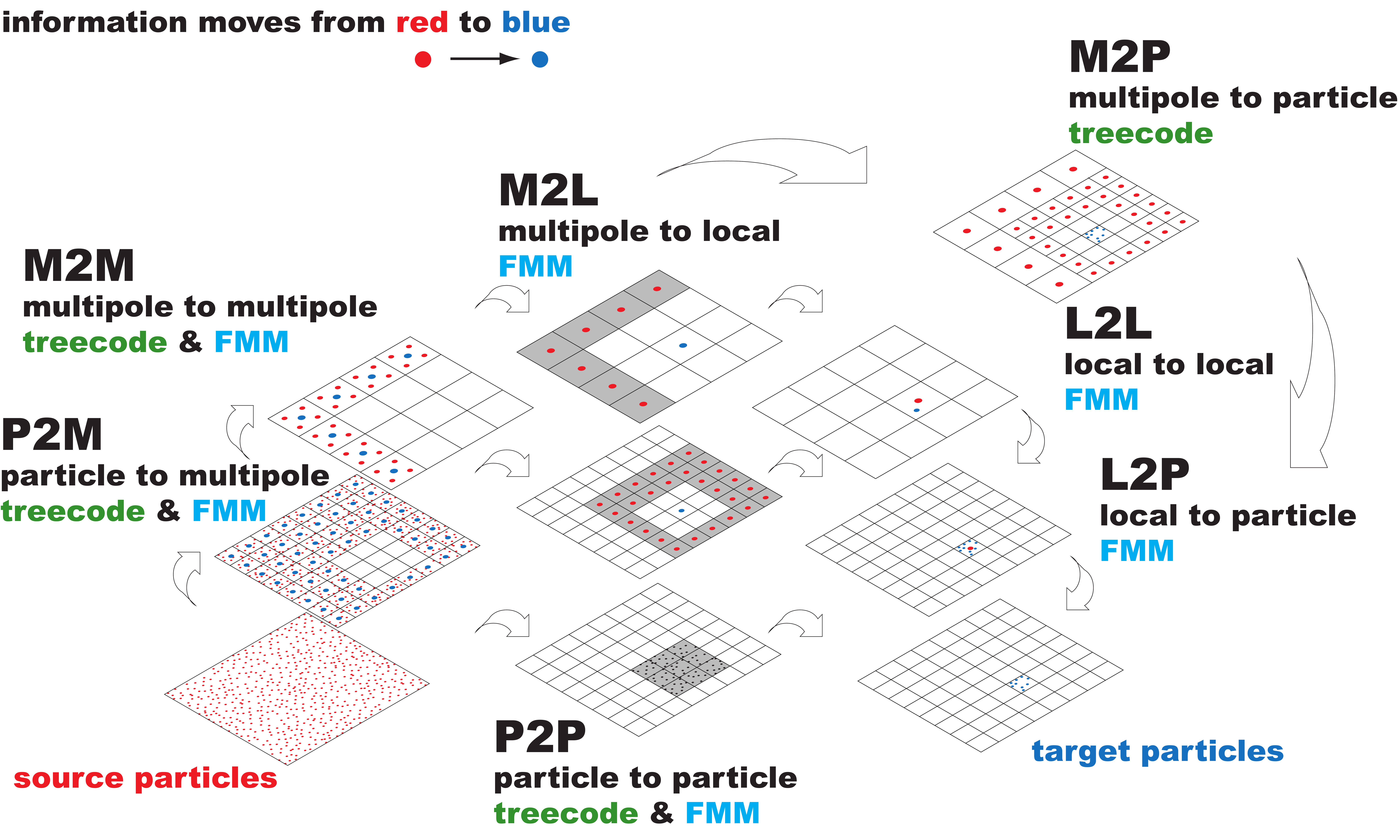}
\end{center}
\caption{Flow of the treecode and FMM calculation.}
\label{fig:flow_chart}
\end{figure}

The flow of the treecode/\fmm calculation is illustrated in Figure \ref{fig:flow_chart}. This schematic shows how the information of all source particles is propagated to a particular set of target particles. The purpose of this figure is to introduce the naming conventions we use for the 7 distinct operations (\PP, \PM, \MM, \MP, \ML, \LL, \LP, \PP), and to associate these steps to a graphical representation. These naming conventions and graphical representations are used later to describe the \gpu\ implementation and to assess its performance. The difference between the treecode and \fmm can be explained concisely using this illustration.

First, the mass/charges of the particles are aggregated into the multipole expansions by calculating $M_{n}^{m}$ at the center of all cells (the \PM\ operation). Next, the multipole expansions are further clustered by translating the center of each expansion to a larger cell and adding their contributions at that level (\MM\ operation). Once the multipole expansions at all levels of the tree are obtained, the treecode calculates Eq.~(\ref{eq:multipole}) to influence the target particles directly (the \MP\ operation). In contrast, the \fmm first transforms the multipole expansions to local expansions (\ML\ operation), and then translates the center of each expansion to smaller cells (\LL\ operation). Finally, the influence of the far field is transmitted from the local expansions to the target particles by calculating Eq.~(\ref{eq:local}) in the \LP\ operation. The influence of the near field is calculated by an all-pairs interaction of neighboring particles (\PP). In the present contribution, all of the above operations are implemented as \gpu\ kernels.

The schematic in Fig.~\ref{fig:flow_chart} shows 2D representations of the actual 3D domain sub-divisions. There are two levels of cell division shown, one with 16 cells and another with 64 cells. For a typical calculation with millions of particles, the tree is further divided into 5 or 6 levels (or more). Recall that the number of cells must be kept proportional to the number of particles for these algorithms to achieve their asymptotic complexity. When there are many levels in the tree, the \MM\ and \LL\ operations are performed multiple times to propagate the information up and down the tree. Also, the \ML\ and \MP\ operations are calculated at every level. The \PM, \LP, and \PP\ are only calculated at the finest (leaf) level of the tree. Since the calculation load decreases exponentially as we move up the tree, the calculation at the leaf level dominates the work load. In particular, it is the \ML/\MP\ and \PP\ that consume most of the runtime in an actual program.

\vspace{\up}
\section{CUDA Implementation of the Fast N-body Algorithms}  
\vspace{\hup}

\begin{figure}[t]
\begin{center}
\includegraphics[width=0.92\textwidth]{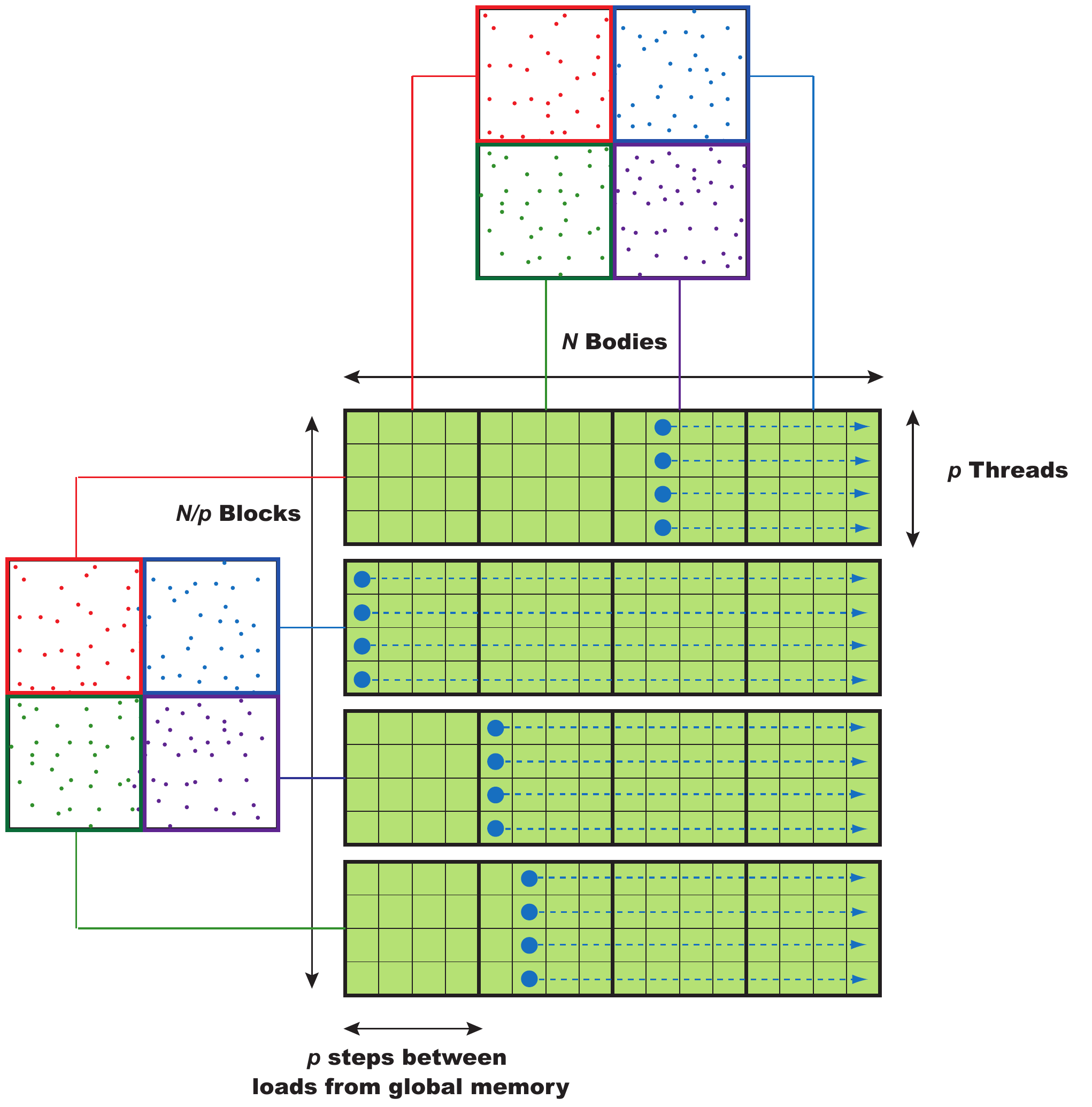}
\end{center}
\caption{Thread block model of the direct evaluation on \gpu; as in \cite{NylandHarrisPrins2007}.}
\label{fig:direct_gpu}
\end{figure}

In our \gpu\ implementation of the treecode and \fmm algorithms we aim for consistency with the  $N$-body example of \cite{NylandHarrisPrins2007}.  Thus, we will utilize their concept of a computational \emph{tile}: a grid consisting of $p$ rows and $p$ columns representing a subset of the pair-wise interactions to be computed.  Consider Fig.~\ref{fig:direct_gpu}, which is adapted from a similar diagram used by the previous authors. Each subset of target particles will be handled by different thread blocks in parallel; the parallel work corresponds to the rows on the diagram. Each subset of source particles is sequentially handled by all thread blocks in chunks of $p$, where $p$ is the number of threads per thread block. As explained in \cite{NylandHarrisPrins2007}: ``Tiles are sized to balance parallelism with data reuse. The degree of parallelism (that is, the number of rows) must be sufficiently large so that multiple warps can be interleaved to hide latencies in the evaluation of interactions. The amount of data reuse grows with the number of columns, and this parameter also governs the size of the transfer of bodies from device memory into shared memory. Finally, the size of the tile also determines the register space and shared memory required."

\begin{figure}
\begin{center}
\includegraphics[width=0.7\textwidth]{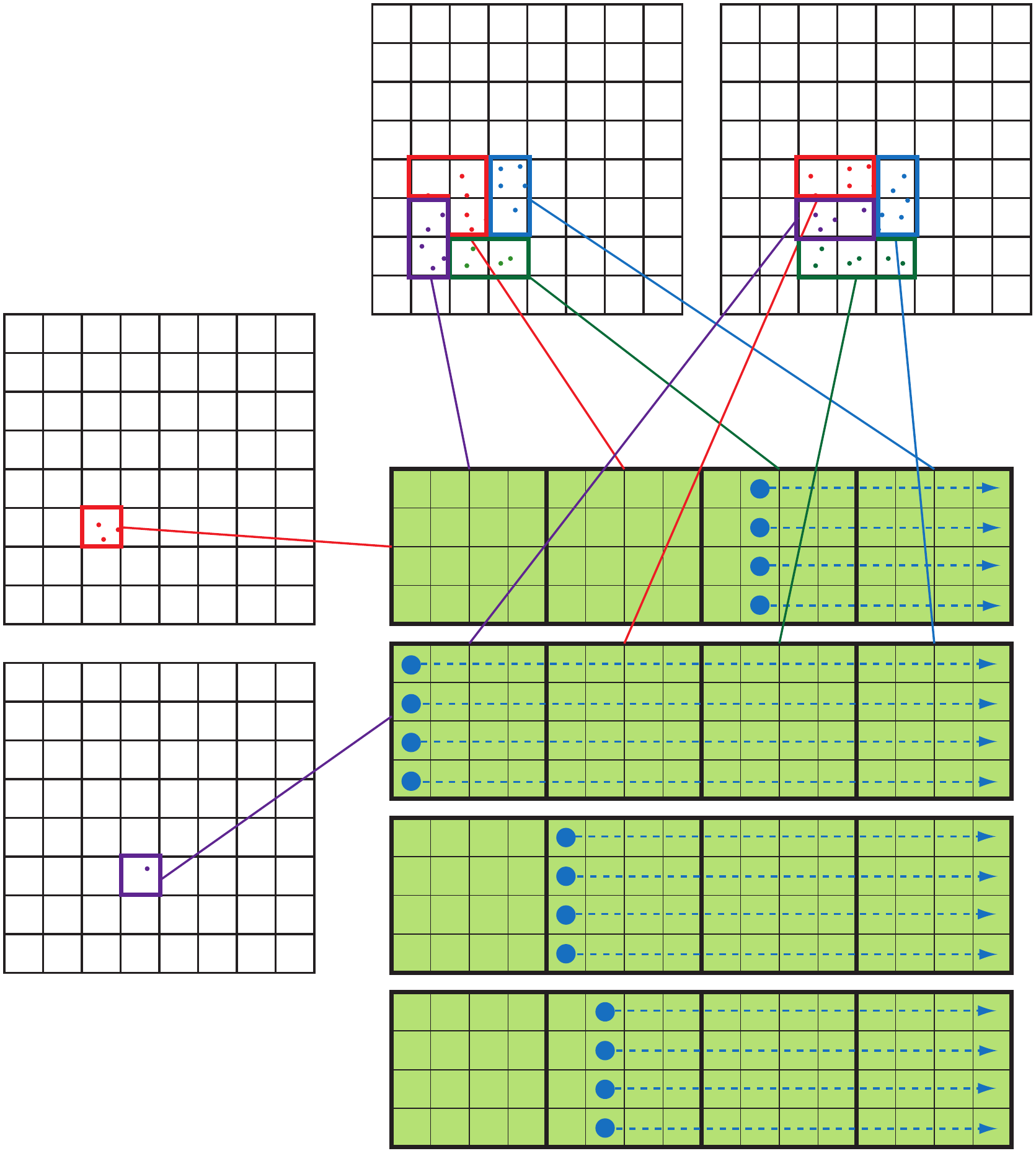}
\end{center}
\caption{Thread block model of the particle-particle interaction on \gpu s.}
\label{fig:p2p_gpu}
\end{figure}

The particle-to-particle (\PP) interactions of the treecode and \fmm are calculated in a similar manner (see Figure \ref{fig:p2p_gpu}). The entire domain is decomposed into an oct-tree, and each cell at the leaf-level is assigned to a thread block. When the number of particles per cell is larger than the size of the thread block, it is split into multiple thread blocks. The main difference with an all-pairs interaction is that each thread block has a different list of source particles. Thus, it is necessary for each thread block to have its unique index list for the offset of source particles. Only the initial offset (for the cells shown in purple in Figure \ref{fig:p2p_gpu}) is passed to the \gpu, and the remaining offsets are determined by increments of $p$.

\begin{figure}
\begin{center}
\includegraphics[width=0.7\textwidth]{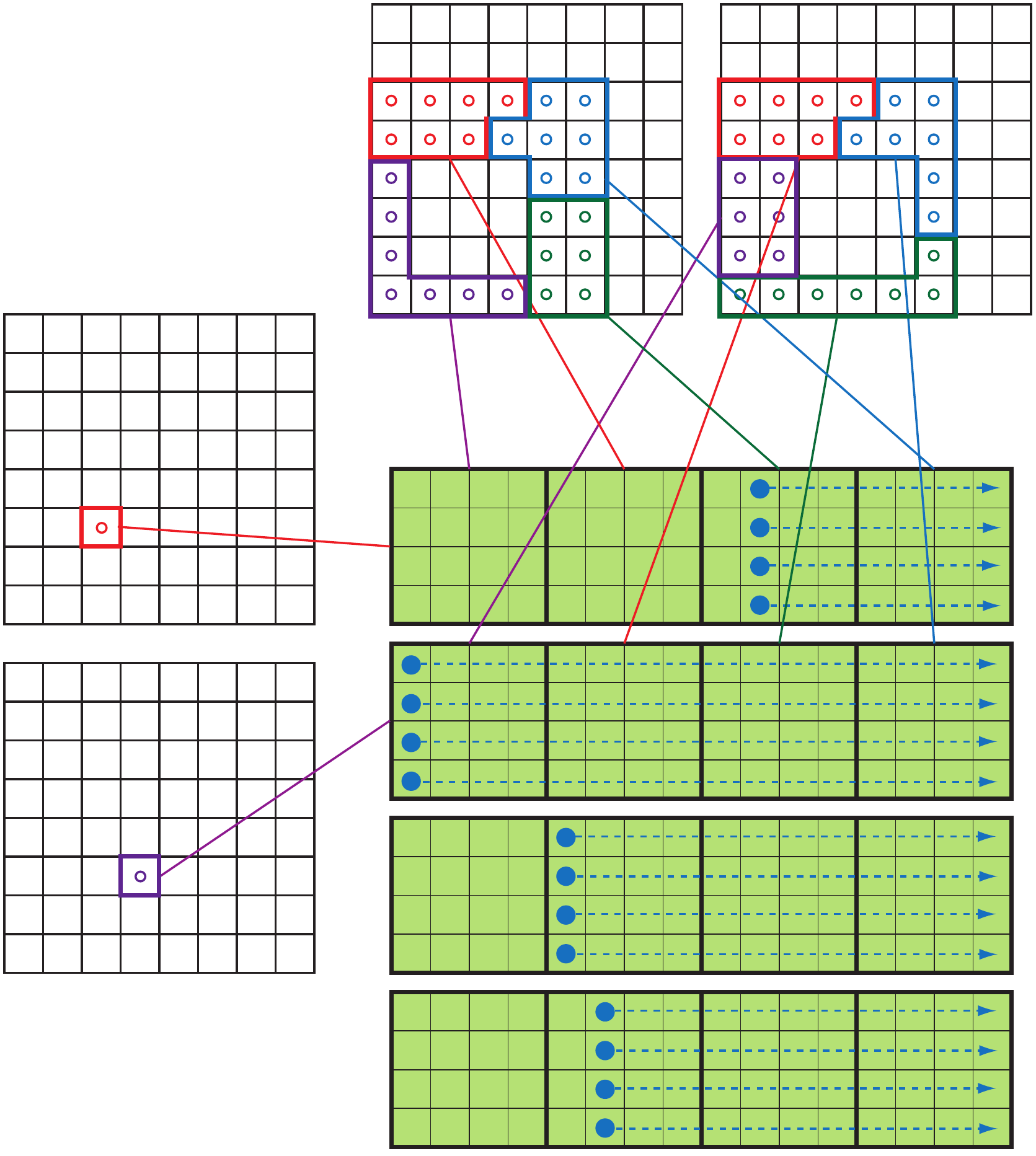}
\end{center}
\caption{Thread block model of the cell-cell interaction on \gpu s}
\label{fig:m2l_gpu}
\end{figure}

In order to ensure coalesced memory access, we accumulate all the source data into a large buffer. On the \cpu, we perform a loop over all interaction lists as if we were performing the actual kernel execution, but instead of calculating the kernel we store the position vector and mass/charge into one large buffer that is passed on to the \gpu. This way, the memory access within the \gpu\ kernel is always contiguous, because the variables are being stored in exactly the same order that they will be accessed. The time it takes to copy the data into the buffer is less than 1\% of the entire calculation. Subsequently, the \gpu\ kernel is called and all the information in the buffer is processed in one call (if it fits in the global memory of the  \gpu). The buffer is split up into an optimum size if it becomes too large to fit on the  global memory. We also create a buffer for the target particles, which contains the position vectors. Once they are passed to the \gpu, the target buffer will be accessed in strides of $p$, assigning one particle to each thread. Since the source particle list is different for each target cell (see Figure \ref{fig:p2p_gpu}), having particles from two different cells in one thread block causes branching of the instruction. We avoid this by padding the target buffer, instead of accumulating the particles in the next cell. For example, if there are 2000 particles per box and the thread block size is 128, the target buffer will be padded with 48 particles so that it uses 16 thread blocks of size 128 ($16\cdot128=2048$) for that cell. In such a case, 1 out of the 16 thread blocks will be doing $37.5\%$ excess work, which is an acceptable trade-off to avoid branching of the instruction within a thread block.

The implementation model used for the \PP\ calculation can be applied to all other steps in the \fmm. An example for the \ML\ translation kernel is shown in Figure \ref{fig:m2l_gpu}. Instead of having particle information in each cell, the cell-cell interactions contain many expansion coefficients per cell. Thus, it is natural to assign one target expansion coefficient to each thread while assigning the cell itself to a thread block. Since the typical number of expansion coefficients is in the order of 10-100, the padding issue discussed in the previous paragraph has greater consequences for this case. In the simplest \cuda implementation that we wish to present in this contribution, we simply reduce the thread block size $p$ to alleviate the problem. In the case of particle-cell interactions (\PM) or cell-particles interactions (\MP, \LP), the same logic is applied where either the target expansion coefficients or target particles are assigned to each thread, and the source expansion coefficients or source particles are read from the source buffer in a coalesced manner and sequentially processed in strides of $p$.

\vspace{\up}

\section{Improvements of Performance}
\vspace{\hup}
\begin{figure}
\begin{center}
\includegraphics[width=0.8\textwidth]{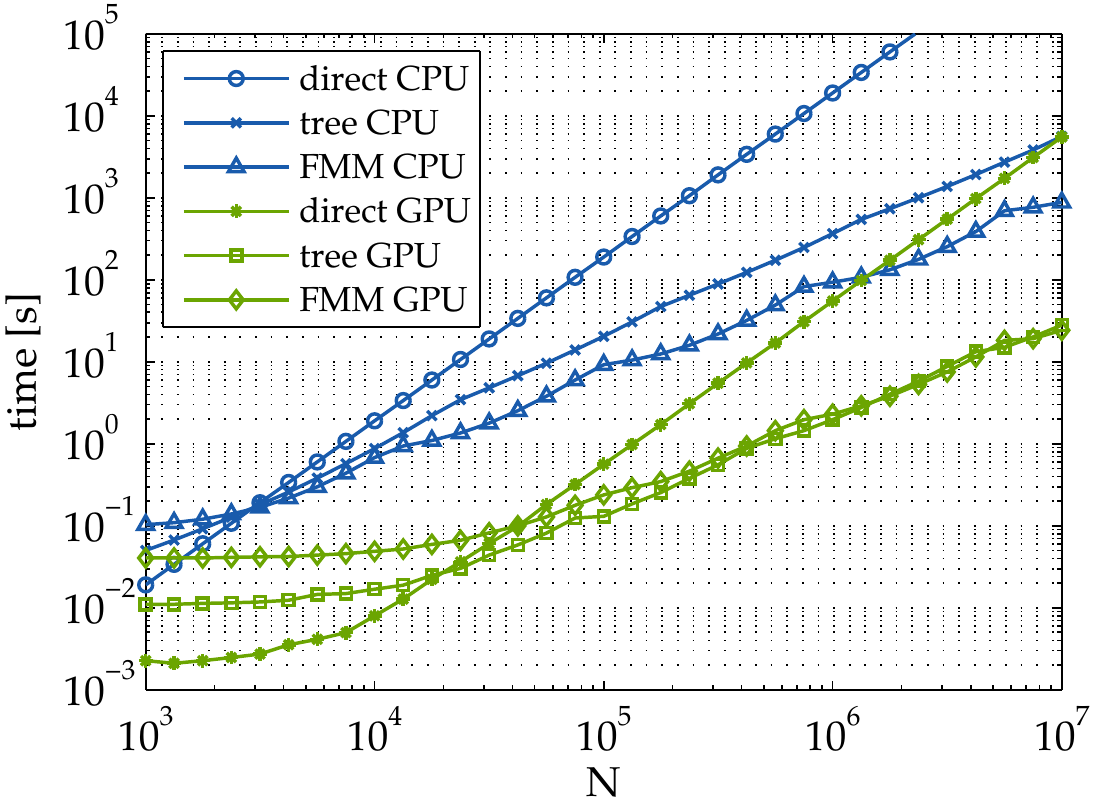}
\end{center}
\caption{Calculation time for the direct method, treecode and \fmm on \cpu\ and \gpu. (Normalized $L^{2}$ norm error of the force is $10^{-4}$ for both treecode and \fmm).}
\label{fig:time_comparision}
\end{figure}

We consider the performance of the treecode and \fmm on \gpu s for the same model problem as in \cite{NylandHarrisPrins2007}.  We would like to point out that the performance metrics shown here apply for the very basic and simplified versions of these kernels. The purpose of this contribution is to show the reader how easy it is to write \cuda programs for the treecode and \fmm. Therefore, many advanced techniques, which would be considered standard for the expert in these algorithms, are deliberately omitted (see section \ref{advanced}). The performance is reported to allow the reader to reproduce the results and verify that their code is performing as expected, and to motivate the discussion about the importance of fast algorithms; we do not claim that the kernels here are as fast as they could be. The \cpu\ tests were run on an Intel Core i7 2.67 GHz, and the \gpu\ tests on an \textsc{NVIDIA 295GTX}. The gcc-4.3 compiler with option -\texttt{O3} was used to compile the \cpu\ codes and \texttt{nvcc} with \texttt{-use\_fast\_math} was used to compile the \cuda codes.

Figure \ref{fig:time_comparision} shows the calculation time against the number of bodies for the direct evaluation, treecode and \fmm on a \cpu\ and \gpu.  The direct calculation is about 300 times faster on the \gpu, compared to the single-core \cpu. The treecode and \fmm are approximately 100 and 30 times faster on the \gpu, respectively. For $N<10^4$, the overhead in the tree construction degrades the performance of the \gpu\ versions. The crossover point between the treecode and direct evaluation is $3\times10^3$ on the \cpu\ and $2\times10^4$ on the \gpu; the crossover point between the \fmm and direct evaluation is $3\times10^3$ on the \cpu\ and $4\times10^4$ on the \gpu. Note that both for the treecode and \fmm, the number of particles at the leaf-level of the tree is higher on the \gpu, to obtain a well-balanced calculation (\emph{i.e.}, comparable time should be spent on the near field and on the far field). The crossover point between the treecode and \fmm is $3\times10^3$ on the \cpu, but is unclear on the \gpu, for the range of our tests.

\begin{figure}
\begin{center}
\includegraphics[width=0.85\textwidth]{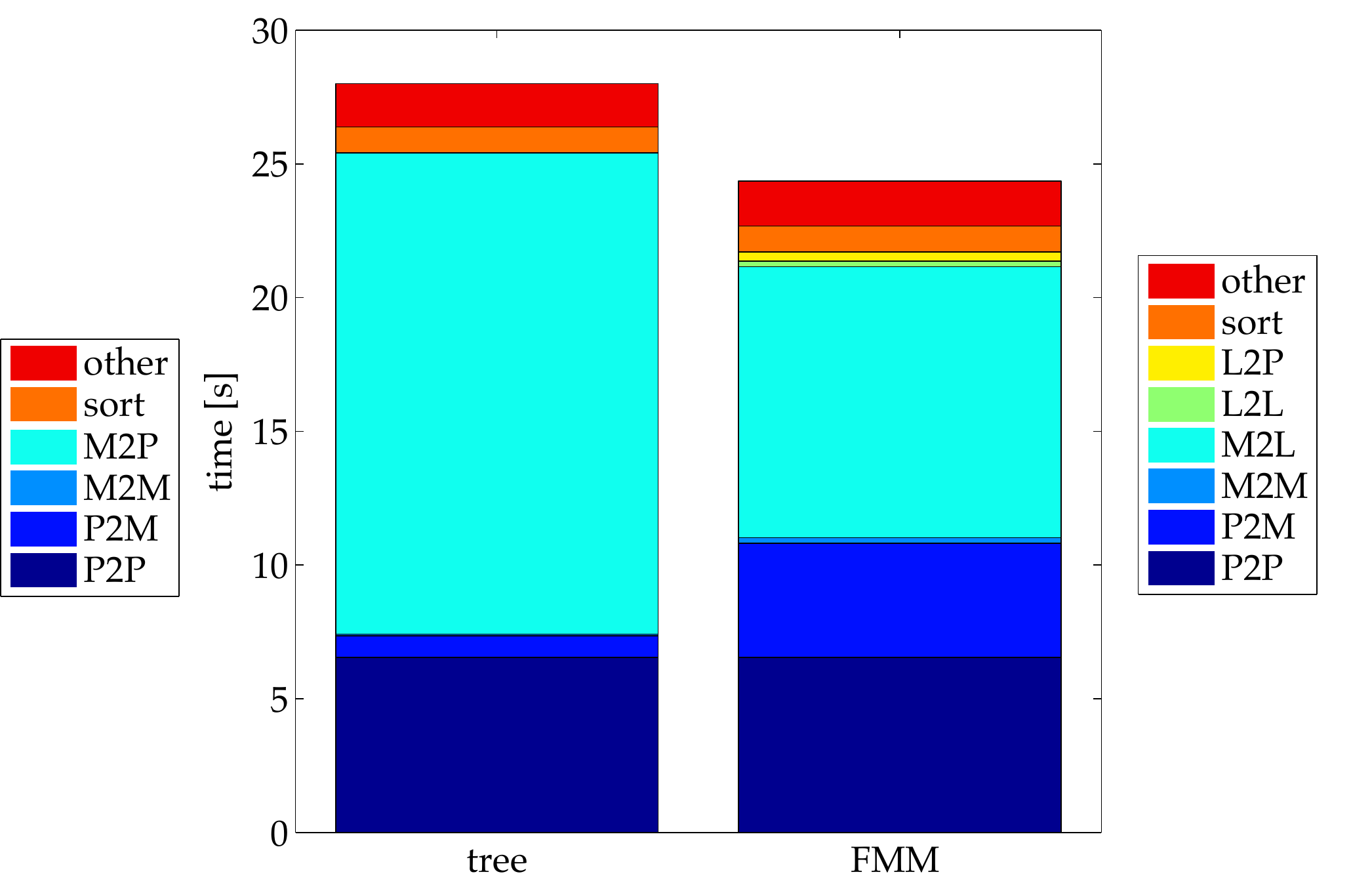}
\end{center}
\caption{Breakdown of the calculation time for the treecode and \fmm on \gpu s using $N=10^7$ particles.}
\label{fig:breakdown}
\end{figure}

When the treecode and \fmm\ are performed on the \cpu, the \PP\ and \MP/\ML\ consume more than 99\% of the execution time. When these computationally-intensive parts are executed on the \gpu, the execution times of the other stages are no longer negligible. This can be seen in the breakdown shown in Figure \ref{fig:breakdown} for the $N=10^7$ case. The contribution of each stage is stacked on top of one another, so the total height of the bar is the total execution time. The legend on the left and right correspond to the treecode and \fmm, respectively; ``sort'' indicates the time it takes to reorder the particles so that they are contiguous within each cell; ``other'' is the total of everything else, including memory allocation, tree construction, interaction list generation, \textit{etc.} The ``sort'' and ``other'' operations are performed on the \cpu. The depth of the tree in this benchmark is the same for both the treecode and \fmm.

As shown in Figure \ref{fig:breakdown}, the \PP\ takes the same amount of time for the treecode and \fmm. This is due to the fact that we use the same neighbor list for the treecode and \fmm. It may be worth noting that the standard treecode uses the distance between particles to determine the clustering threshold (for a given desired accuracy), and has an interaction list that is slightly more flexible than that of the \fmm. A common measure to determine the clustering in treecodes is the Barnes-Hut multiple acceptance criteria (\mac) $\theta>l/d$ \cite{BarnesHut1986}, where $l$ is the size of the cell, and $d$ is the distance between the particle and center of mass of the cell. The present calculation uses the standard \fmm\ neighbor list shown in Figure \ref{fig:flow_chart} for both the \fmm\ and treecode, which results in a  \mac\ of $\theta=2/3$.
The \PM\ operation takes longer for the \fmm\ because the order of multipole expansions is larger than in the treecode, to achieve the same accuracy. The calculation loads of \MM, \LL\ and \LP\ are small compared to the \MP\ and \ML. The \MP\ has a much larger calculation load than the \ML, but it has more data-parallelism. Therefore, the \gpu\ implementation of these two kernels has a somewhat similar execution time. The high data-parallelism of the \MP\ is an important factor we must consider when comparing the treecode and \fmm\ on \gpu s.

\begin{figure}
\begin{center}
\includegraphics[width=0.7\textwidth]{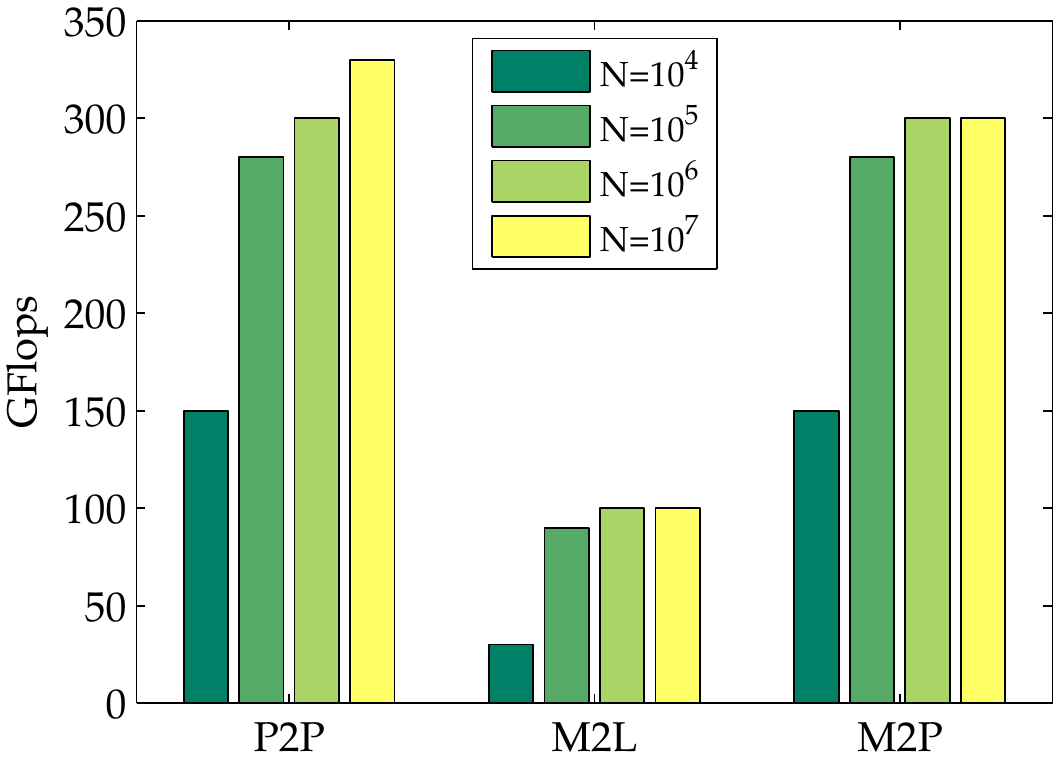}
\end{center}
\caption{Actual performance in Gflop/s of three core kernels, for different values of $N$.}
\label{fig:gflops}
\end{figure}

Figure \ref{fig:gflops} shows the measured performance on the \gpu\, measured in Gflop/s; this is actual operations performed in the code, \emph{i.e.}, a \texttt{sqrt} counts 1, etc.  Clearly, for $N=10^{4}$ the \gpu\ is underutilized, but performance is quite good for the larger values of $N$.  The \PP\ operation performs very well, achieving in the order of 300 Gflop/s for the larger values of $N$ of these tests.  The \MP\ performs much better than the \ML, due to the higher inherent parallelism.  This explains why we see the treecode accelerating better overall, compared to \fmm, on Figure \ref{fig:time_comparision}.
\vspace{-6mm}
\section{Detailed description of the \gpu\ kernels}
\vspace{\hup}

In this section, we give a detailed explanation of the implementation of the treecode/\fmm\ in \cuda. The code snippets shown here are extracted directly from the code available from the distribution released with this article\footnote{All source code can be found in \href{http://code.google.com/p/gemsfmm/}{\textsf{http://code.google.com/p/gemsfmm/}}}. In particular, we will describe the implementation of the \PP\ and \ML\ kernels, which take up most of the calculation time.

\vspace{\hup}
\subsection{The \PP\ kernel implementation}
\vspace{\hup}

We start with the simplest kernel for the interaction of a single pair of particles, shown in Listing \ref{lst:p2p_kernel_core}. Equation (\ref{eq:force}) is calculated here without the $m_i$. In other words, it is the acceleration $a_i=F_i/m_i$ that is being calculated. This part of the code is very similar to that of the {\ttfamily\small nbody} example in the \cuda\ \sdk, which is explained in detail in \cite{NylandHarrisPrins2007}. The only difference is that the present kernel uses the reciprocal square-root function instead of a square-root and division. There are 19 floating-point operations in this kernel, counting the 3 additions, 6 subtractions, 9 multiplications, and 1 reciprocal square-root. The list of variables is as follows:

\vspace{\up}
\begin{itemize}

\item {\ttfamily\small posTarget} is the position vector of the target particles; it has a {\ttfamily\small float3} data type and is stored in registers.\vspace{\hup}

\item {\ttfamily\small sharedPosSource} is the position vector and the mass of the source particles; it has a {\ttfamily\small float4} data type and resides in shared memory.\vspace{\hup}

\item {\ttfamily\small accel} is the acceleration vector of the target particles; it has a {\ttfamily\small float3} data type and is stored in registers.\vspace{\hup}

\item  the {\ttfamily\small float3} data type is used to store the distance vectors {\ttfamily\small dist}.\vspace{\hup}

\item {\ttfamily\small eps} is the softening factor \cite[see][]{Aarseth2003}.\vspace{\hup}
\end{itemize}

\begin{lstlisting}[float=t,frame=tb,caption=\PP\ kernel for a single interaction,label=lst:p2p_kernel_core]
__device__ float3 p2p_kernel_core(float3 accel,
                                  float3 posTarget, float4 sharedPosSource)
{
  float3 dist;
  dist.x = posTarget.x - sharedPosSource.x;
  dist.y = posTarget.y - sharedPosSource.y;
  dist.z = posTarget.z - sharedPosSource.z;
  float invDist = rsqrtf(dist.x * dist.x + dist.y * dist.y + dist.z * dist.z + eps);
  float invDistCube = invDist * invDist * invDist;
  float s = sharedPosSource.w * invDistCube;
  accel.x  -=  dist.x * s;
  accel.y  -=  dist.y * s;
  accel.z  -=  dist.z * s;
  return accel;
}
\end{lstlisting}

The function shown in Listing \ref{lst:p2p_kernel_core} is called from an outer kernel which calculates the pairwise interactions of all particles in the \PP\ interaction list. This outer kernel is shown in Listing \ref{lst:p2p_kernel}, and its graphical representation is shown in Figure \ref{fig:p2p_gpu}. The input variables are {\ttfamily\small deviceOffset}, {\ttfamily\small devicePosTarget}, {\ttfamily\small devicePosSource}, and the output is {\ttfamily\small deviceAccel}. The description of these variables is as follows:

\vspace{\up}
\begin{itemize}
\item {\ttfamily\small deviceOffset} contains the number of interacting cells and the offset of the particle index for each of these cells;\vspace{\hup}

\item  {\ttfamily\small devicePosTarget} contains the position vector of the target particles;\vspace{\hup}

\item   {\ttfamily\small devicePosSource} is the position vector of the source particles, and \vspace{\hup}

\item {\ttfamily\small deviceAccel} is the acceleration vector of target particles.\vspace{\hup}

\end{itemize}

\begin{lstlisting}[float=t!,frame=tb,caption=The entire \PP\ kernel,label=lst:p2p_kernel]
__global__ void p2p_kernel(int* deviceOffset, float3* devicePosTarget,
                           float4* devicePosSource, float3* deviceAccel)
{
  int jbase, jsize, jblok, numInteraction;
  int j, ij, jj, jb;
  const int threadsPerBlock = threadsPerBlockTypeA;
  const int offsetStride = 2 * maxP2PInteraction + 1;
  float3 posTarget;
  float3 accel = {0.0f, 0.0f, 0.0f};
  __shared__ float4 sharedPosSource[threadsPerBlock];
  posTarget = devicePosTarget[blockIdx.x * threadsPerBlock + threadIdx.x];
  numInteraction = deviceOffset[blockIdx.x * offsetStride];
  for(ij = 0; ij < numInteraction; ij++){
    jbase = deviceOffset[blockIdx.x * offsetStride + 2 * ij + 1];
    jsize = deviceOffset[blockIdx.x * offsetStride + 2 * ij + 2];
    jblok = (jsize + threadsPerBlock - 1) / threadsPerBlock;
    for(j = 0; j < jblok-1; j++){
      jb = jbase + j * threadsPerBlock + threadIdx.x;
      sharedPosSource[threadIdx.x] = devicePosSource[jb];
      __syncthreads();
#pragma unroll 32
      for(jj = 0; jj < threadsPerBlock; jj++){
        accel = p2p_kernel_core(accel, posTarget, sharedPosSource[jj]);
      }
      __syncthreads();
    }
    jb = jbase + j * threadsPerBlock + threadIdx.x;
    sharedPosSource[threadIdx.x] = devicePosSource[jb];
    __syncthreads();
    for(jj = 0; jj < jsize - (j * threadsPerBlock); jj++){
      accel = p2p_kernel_core(accel, posTarget, sharedPosSource[jj]);
    }
    __syncthreads();
  }
  deviceAccel[blockIdx.x * threadsPerBlock + threadIdx.x] = accel;
}
\end{lstlisting}

 All variables that begin with ``{\ttfamily\small device}" are stored in the device memory. All variables that begin with ``{\ttfamily\small shared}" are stored in shared memory. Everything else is stored in the registers. Lines 4--10 are declaration of variables; it is possible to reduce register space usage by reusing some of these variables, but for pedagogical purposes we have chosen to declare each variable that has a different functionality. There are 4 variables that are defined externally. One is the {\ttfamily\small threadsPerBlockTypeA}, which is the number of threads per thread-block for the \PP\ kernel. We use a different number of threads per thread-block, {\ttfamily\small threadsPerBlockTypeB}, for the other kernels that have expansion coefficients as targets. On line 5, {\ttfamily\small threadsPerBlockTypeA} is passed to {\ttfamily\small threadsPerBlock} as a constant. Another external variable is used on line 7, where {\ttfamily\small maxP2PInteraction} (the maximum number of neighbor cells in a \PP\ interaction) is used to calculate {\ttfamily\small offsetStride} (the stride of the data in {\ttfamily\small deviceOffset}). The other two externally defined variables are {\ttfamily\small threadIdx} and {\ttfamily\small blockIdx}, which are thread index and thread-block index provided by \cuda.

On line 11, the position vectors are copied from the global memory to the registers. On line 12, the number of interacting cells is read from the {\ttfamily\small deviceOffset}, and on line 13 this number is used to form a loop that goes through all the interacting cells (27 cells for the \PP\ interaction). Note that each thread block handles (part of) only one target cell, and the interaction list of the neighboring cells is identical for all threads within the thread block. In other words, {\ttfamily\small blockIdx.x} identifies which target cell we are looking at, and {\ttfamily\small ij} identifies which source cell it is interacting with. On line 14, the offset of the particle index for that source cell is copied from {\ttfamily\small deviceOffset} to {\ttfamily\small jbase}. On line 15, the number of particles in the source cell is copied to {\ttfamily\small jsize}.
Now we have the information of the target particles and the offset and size of the source particles that they interact with. At this point, the information of the source particles still resides in the device memory. This information is copied to the shared memory in coalesced chunks of size {\ttfamily\small threadsPerBlock}. However, the number of particles per cell is not always a multiple of {\ttfamily\small threadsPerBlock}, so the last chunk will contain a remainder that is different from {\ttfamily\small threadsPerBlock}. It is inefficient to have a conditional branching to detect if the chunk is the last one or not, and it is a waste of storage to pad for each source cell. Therefore, on line 16 the number of chunks {\ttfamily\small jblok} is calculated by rounding up {\ttfamily\small jsize} to the nearest multiple of {\ttfamily\small threadsPerBlock}. On line 17, a loop is executed for all chunks except the last one. The last chunk is processed separately on lines 27--33. On line 18, the index of the source particle on the device memory is calculated by offsetting the thread index first by the chunk offset {\ttfamily\small j*threadsPerBlock} and then by the cell offset {\ttfamily\small jbase}. On line 19, this global index is used to copy the position vector of the source particles from device memory to shared memory. Subsequently, {\ttfamily\small \_\_syncthreads()} is called to ensure that the copy to shared memory has completed on all threads before proceeding. On lines 21--24, a loop is performed for all elements in the current chunk of source particles, where the {\ttfamily\small p2p\_kernel\_core} is called per pairwise interaction. The {\ttfamily\small \#pragma unroll 32} is the same loop unrolling suggested in \cite{NylandHarrisPrins2007}. On line 25, {\ttfamily\small \_\_syncthreads()} is called to keep {\ttfamily\small sharedPosSource} from being overwritten for the next chunk before having been used in the current one. Lines 27--33 are identical to lines 18--25 except for the loop counter for {\ttfamily\small jj}, which is the remainder instead of {\ttfamily\small threadsPerBlock}. On line 35, the acceleration vector in registers is copied back to the device memory by offsetting the thread index by {\ttfamily\small blockIdx.x * threadsPerBlock}.

\vspace{\hup}
\subsection{The \ML\ kernel implementation}
\vspace{\hup}

As shown in Equations (\ref{eq:multipole}) and (\ref{eq:local}), the multipole-to-local translation in the \fmm\ is the translation of the multipole expansion coefficients $M_{n}^{m}$ in one location to the local expansion coefficients $L_{n}^{m}$ at another. If we relabel the indices of the local expansion matrix to $L_{j}^{k}$, the \ML\ translation can be written as
\begin{equation}
L^{k}_{j}=\sum^{p-1}_{n=0}\sum^{n}_{m=-n}\frac{M^{m}_{n}i^{|k-m|-|k|-|m|}A^{m}_{n}A^{k}_{j}Y^{m-k}_{j+n}(\alpha,\beta)}{(-1)^{j}A^{m-k}_{j+n}\rho^{j+n+1}}
\label{eq:m2l}
\end{equation}
where $i$ is the imaginary unit, $p$ is the order of the series expansion, $A_{n}^{m}$ is defined as
\begin{equation}
A^{m}_{n}=\frac{1}{\sqrt{(n-m)!(n+m)!}}
\label{eq:anm}
\end{equation}
and $Y_{n}^{m}$ is the spherical harmonic
\begin{equation}
Y^{m}_{n}(\alpha,\beta)=\sqrt{\frac{(n-\left|m\right|)!}{(n+\left|m\right|)!}}P^{\left|m\right|}_{n}(\cos\alpha)e^{im\beta}.
\label{eq:ynm}
\end{equation}

In order to calculate the spherical harmonics, the value of the associated Legendre polynomials $P^{m}_{n}$ must be determined. The associated Legendre polynomials have a recurrence relation, which require only the information of $x=\cos\alpha$ to start. The recurrence relations and identities used to generate the full associated Legendre polynomial are,
\begin{align}
(n-m+1)P^{m}_{n+1}(x)&=x(2n+1)P^{m}_{n}(x)-(n+m)P^{m}_{n-1}(x),\label{eq:pnm1}\\
P^{m}_{m}(x)&=(-1)^m(2m-1)!(1-x^2)^{m/2},\label{eq:pnm2}\\
P^{m}_{m+1}&=x(2m+1)P^{m}_{m}(x)\label{eq:pnm3}
\end{align}

The \ML\ kernel calculates Equation (\ref{eq:m2l}) in two stages. First, $Y_{n}^{m}/\rho^{n+1}/A_{n}^{m}$ is calculated using Equations (\ref{eq:anm})--(\ref{eq:pnm3}). Then, Equation (\ref{eq:m2l}) is calculated by substituting this result after switching the indices $n\rightarrow j+n$ and $m\rightarrow m-k$. Thus, $M_{n}^{m}i^{|k-m|-|k|-|m|}A_{n}^{m}A_{j}^{k}/(-1)^{j}$ is calculated at the second stage. Furthermore, in the \gpu\ implementation the complex part $e^{im\beta}$ in Equation (\ref{eq:ynm}) is multiplied at the end of the second stage so that the values remain real until then. At the end of the second stage, we simply put the real and complex part of the $L_{j}^{k}$ into two separate variables.

\begin{lstlisting}[float=t!,frame=tb,caption=Calculation of the spherical harmonic for the \ML\ kernel,label=lst:m2l_calculate_ynm]
__device__ void m2l_calculate_ynm(float* sharedYnm,
                                  float rho, float alpha, float* sharedFactorial)
{
  int i, m, n;
  float x, s, fact, pn, p, p1, p2, rhom, rhon;
  x = cosf(alpha);
  s = sqrt(1 - x * x);
  fact = 1;
  pn = 1;
  rhom = 1.0 / rho;
  for(m = 0; m < 2 * numExpansions; m++){
    p = pn;
    i = m * (m + 1) /2 + m;
    sharedYnm[i] = rhom * p;
    p1 = p;
    p = x * (2 * m + 1) * p;
    rhom /= rho;
    rhon = rhom;
    for(n = m + 1; n < 2 * numExpansions; n++){
      i = n * (n + 1) / 2 + m;
      sharedYnm[i] = rhon * p * sharedFactorial[n - m];
      p2 = p1;
      p1 = p;
      p = (x * (2 * n + 1) * p1 - (n + m) * p2) / (n - m + 1);
      rhon /= rho;
    }
    pn = -pn * fact * s;
    fact = fact + 2;
  }
}
\end{lstlisting}

The \gpu\ implementation of the first part for $Y_{n}^{m}/\rho^{n+1}/A_{n}^{m}$ is shown in Listing \ref{lst:m2l_calculate_ynm}. As was the case with Listing \ref{lst:p2p_kernel_core}, this function is called from an outer function that calculates the entire \ML\ translation for all cells. The inputs are {\ttfamily\small rho}, {\ttfamily\small alpha}, and {\ttfamily\small sharedFactorial}. The output is {\ttfamily\small sharedYnm}. Since, we do not calculate the $e^{im\beta}$ part of the spherical harmonic at this point, {\ttfamily\small beta} is not necessary. {\ttfamily\small sharedFactorial} contains the values of the factorials for a given index, \textit{i.e.} {\ttfamily\small sharedFactorial[n]}$=n!$. Also, it is $Y_{n}^{m}/\rho^{n+1}/A_{n}^{m}$ that is stored in {\ttfamily\small sharedYnm} and not $Y_{n}^{m}$ itself. Basically, Equation (\ref{eq:pnm1}) is calculated on line 24, Equation (\ref{eq:pnm2}) is calculated on line 27, and Equation (\ref{eq:pnm3}) is calculated on line 16. {\ttfamily\small p}, {\ttfamily\small p1}, and {\ttfamily\small p2} correspond to $P_{n+1}^{m}$, $P_{n}^{m}$, and $P_{n-1}^{m}$, respectively. However, {\ttfamily\small p} is used in lines 14 and 21 before it is updated on lines 16 and 24, so it represents $P_{n}^{m}$ at the time of usage. This $P_{n}^{m}$ is used to calculate $Y_{n}^{m}/\rho^{n+1}/A_{n}^{m}$ on lines 14 and 21, although the correspondence to the equation is not obvious at first hand. The connection to the equation will become clear when we do the following transformation,
\begin{equation}
\frac{Y_{n}^{m}}{\rho^{n+1}A_{n}^{m}}=\frac{\sqrt{(n-m)!/(n+m)!}P^{m}_{n}e^{im\beta}}{\rho^{n+1}/\sqrt{(n-m)!(n+m)!}}=\frac{(n-m)!P_{n}^{m}}{\rho^{n+1}}e^{im\beta}
\end{equation}

As mentioned earlier, we do not calculate the $e^{im\beta}$ at this point so {\ttfamily\small sharedYnm} is symmetric with respect to the sign of $m$. Therefore, the present loop for the recurrence relation is performed for only $m\ge0$ and the absolute sign for $m$ in Equation (\ref{eq:ynm}) disappears. We can also save shared memory consumption by storing only the $m\ge0$ half of the spherical harmonic in {\ttfamily\small sharedYnm}.

The second stage of the \ML\ kernel is shown in Listing \ref{lst:m2l_kernel_core}. The inputs are {\ttfamily\small j}, {\ttfamily\small beta}, {\ttfamily\small sharedFactorial}, {\ttfamily\small sharedYnm}, and {\ttfamily\small sharedMnmSource}. The output is {\ttfamily\small LnmTarget}. In this second stage of the \ML, the remaining parts of Equation (\ref{eq:m2l}) are calculated to obtain $L_{j}^{k}$. Each thread handles a different coefficient in $L_{j}^{k}$. In order to do this, we must associate the {\ttfamily\small threadIdx.x} to a pair of  {\ttfamily\small j} and  {\ttfamily\small k}. In the outer function, which will be shown later, the index {\ttfamily\small j} corresponding to {\ttfamily\small threadIdx.x} is calculated and passed to the present function. Lines 9--11, determine the index {\ttfamily\small k} from the input {\ttfamily\small j} and {\ttfamily\small threadIdx.x}.

\begin{lstlisting}[float=t!,frame=tb,caption=Calculation of $L_{n}^{m}$ in the \ML\ kernel,label=lst:m2l_kernel_core]
__device__ void m2l_kernel_core(float* LnmTarget,
                                int j, float beta,
                                float* sharedFactorial,
                                float* sharedYnm,
                                float* sharedMnmSource)
{
  int i, k, m, n, jnkm;
  float ere, eim, anm, ajk, cnm, CnmReal, CnmImag;
  k = 0;
  for(i = 0; i <= j; i++) k += i;
  k = threadIdx.x - k;
  // using pre-processed function ODDEVEN
  ajk = ODDEVEN(j) * rsqrtf(sharedFactorial[j - k] * sharedFactorial[j + k]);
  for(n = 0; n < numExpansions; n++){
    for(m = -n; m < 0; m++){
      i = n * (n + 1) / 2 - m;
      jnkm = (j + n) * (j + n + 1) / 2 - m + k;
      ere = cosf((m - k) * beta);
      eim = sinf((m - k) * beta);
      anm = rsqrtf(sharedFactorial[n - m] * sharedFactorial[n + m]);
      cnm = anm * ajk * sharedYnm[jnkm];
      CnmReal = cnm * ere;
      CnmImag = cnm * eim;
      LnmTarget[0] += sharedMnmSource[2 * i + 0] * CnmReal;
      LnmTarget[0] += sharedMnmSource[2 * i + 1] * CnmImag;
      LnmTarget[1] += sharedMnmSource[2 * i + 0] * CnmImag;
      LnmTarget[1] -= sharedMnmSource[2 * i + 1] * CnmReal;
    }
    for(m = 0; m <= n; m++){
      i = n * (n + 1) / 2 + m;
      jnkm = (j + n) * (j + n + 1) / 2 + abs(m - k);
      ere = cosf((m - k) * beta);
      eim = sinf((m - k) * beta);
      anm = rsqrtf(sharedFactorial[n - m] * sharedFactorial[n + m]);
      cnm = ODDEVEN((abs(k - m) - k - m) / 2);
      cnm *= anm * ajk * sharedYnm[jnkm];
      CnmReal = cnm * ere;
      CnmImag = cnm * eim;
      LnmTarget[0] += sharedMnmSource[2 * i + 0] * CnmReal;
      LnmTarget[0] -= sharedMnmSource[2 * i + 1] * CnmImag;
      LnmTarget[1] += sharedMnmSource[2 * i + 0] * CnmImag;
      LnmTarget[1] += sharedMnmSource[2 * i + 1] * CnmReal;
    }
  }
}

\end{lstlisting}

We will remind the reader again that this part of the \ML\ kernel calculates $M_{n}^{m}i^{|k-m|-|k|-|m|}$ $A_{n}^{m}A_{j}^{k}/(-1)^{j}$. This results in a quadruple loop over the indices $j$, $k$, $m$, and $n$. However, in the \gpu\ implementation the first two indices are thread-parallelized, only leaving $m$ and $n$ as sequential loops starting from lines 13, 14, and 28. Lines 14--27 are for negative $m$, while lines 28--42 are for positive $m$. $A_{j}^{k}/(-1)^{j}$ is calculated on line 12. We define a preprocessed function ``{\ttfamily\small \#define ODDEVEN(n) ((n \& 1  ==  1) ? -1 : 1)}", which calculates $(-1)^n$ without using a power function. $A_{n}^{m}$ is calculated on lines 19 and 33. $i^{|k-m|-|k|-|m|}$ is calculated on line 34 for the $m\ge0$ case, and is always $1$ for $m<0$. Since $|k-m|-|k|-|m|$ is always an even number, it is possible to calculate $i^{|k-m|-|k|-|m|}$ as $-1^{(|k-m|-|k|-|m|)/2}$ and use the {\ttfamily\small ODDEVEN} function defined previously.
Then, {\ttfamily\small anm}, {\ttfamily\small ajk}, and {\ttfamily\small sharedYnm} are multiplied to this result. The complex part $e^{im\beta}$ that was omitted in the first stage is calculated on lines 17--18 and 31--32 using the index $m-k$ instead of $m$; {\ttfamily\small ere} is the real part and {\ttfamily\small eim} is the imaginary part. {\ttfamily\small CnmReal} and {\ttfamily\small CnmImag} in lines 21--22 and 36--37 are the real and imaginary parts of the product of all the terms described above. Finally, these values are multiplied to $M_{n}^{m}$ in lines 23--26 and 38--41, where {\ttfamily\small sharedMnmSource[2*i+0]} is the real part and {\ttfamily\small sharedMnmSource[2*i+1]} is the imaginary part. We use the relation $M_{n}^{-m}=\overline{M_{n}^{m}}$ to reduce the storage of {\ttfamily\small sharedMnmSource}. Therefore, the imaginary part has opposite signs for the $m\ge0$ case and $m<0$ case. The real part of $L_{j}^{k}$ is accumulated in {\ttfamily\small LnmTarget[0]}, while the imaginary part is accumulated in {\ttfamily\small LnmTarget[1]}.

The functions in Listings \ref{lst:m2l_calculate_ynm} and \ref{lst:m2l_kernel_core} are called from an outer function shown in Listing \ref{lst:m2l_kernel}. This function is similar to the one shown in Listing \ref{lst:p2p_kernel}. The inputs are {\ttfamily\small deviceOffset} and {\ttfamily\small deviceMnmSource}. The output is {\ttfamily\small deviceLnmTarget}.  The definitions are:

\vspace{\up}
\begin{itemize}
\item  {\ttfamily\small deviceOffset} contains the number of interacting cells, the offset of the particle index for each of these cells, and the 3D index of their relative positioning.\vspace{\hup}

\item   {\ttfamily\small threadsPerBlockTypeB} and {\ttfamily\small maxM2LInteraction} are defined externally.\vspace{\hup}

\item   {\ttfamily\small maxM2LInteraction} is the maximum size of the interaction list for the \ML, which is 189 for the present kernels. \vspace{\hup}

\item {\ttfamily\small offsetStride}, calculated on line 6, is the stride of the data in {\ttfamily\small deviceOffset}.\vspace{\hup}
\end{itemize}

On line 8, the size of the cell is read from {\ttfamily\small deviceConstant[0]}, which resides in constant memory. On line 10, {\ttfamily\small LnmTarget} is initialized. Each thread handles a different coefficient in $L_{j}^{k}$. In order to do this, we must associate the {\ttfamily\small threadIdx.x} to a pair of  {\ttfamily\small j} and  {\ttfamily\small k}. {\ttfamily\small sharedJ} returns the index $j$ when given the {\ttfamily\small threadIdx.x} as input. It is declared on line 11, initialized on lines 16--18, the values are calculated on lines 19--24, and then passed to {\ttfamily\small m2l\_kernel\_core()} on line 40. {\ttfamily\small sharedMnmSource} is the copy of {\ttfamily\small deviceMnmSource} in shared memory. It is declared on line 12 and the values are copied on lines 35--36 before it is passed to {\ttfamily\small m2l\_kernel\_core()} on line 41. {\ttfamily\small sharedYnm} contains the real spherical harmonics. It is declared on line 13 and its values are calculated in the function {\ttfamily\small m2l\_calculate\_ynm} on line 39 before they are passed to {\ttfamily\small m2l\_kernel\_core} on line 41. {\ttfamily\small sharedFactorial} contains the factorial for the given index and is declared on line 14 and its values are calculated on lines 25--29 before they are passed to {\ttfamily\small m2l\_kernel\_core} on line 41. On line 15, the number of interacting cells is read from {\ttfamily\small deviceOffset} and its value {\ttfamily\small numInteraction} is used for the loop on line 30. The offset of particles are read from {\ttfamily\small deviceOffset} on line 31, and the relative distance of the source and target cell are calculated on lines 32--34. On line 38, this distance is transformed into spherical coordinates using an externally defined function {\ttfamily\small cart2sph}. The two functions shown in Listings \ref{lst:m2l_calculate_ynm} and \ref{lst:m2l_kernel_core} are called on lines 39--41. Finally, the results in {\ttfamily\small LnmTarget} are copied to {\ttfamily\small deviceLnmTarget} on line 45.

Listings \ref{lst:p2p_kernel_core}--\ref{lst:m2l_kernel} are the core components of the present \gpu\ implementation. We hope that the other parts of the open-source code that we provide along with this article are understandable to the reader without explanation.

\begin{lstlisting}[float=t!,frame=tb,caption=The entire \ML\ kernel,label=lst:m2l_kernel]
__global__ void m2l_kernel(int* deviceOffset, float* deviceLnmTarget,
                           float* deviceMnmSource)
{
  int i, j, k, ij, ib, numInteraction, jbase;
  const int threadsPerBlock = threadsPerBlockTypeB;
  const int offsetStride = 4*maxM2LInteraction+1;
  float3 dist;
  float boxSize = deviceConstant[0];
  float rho, alpha, beta, fact;
  float LnmTarget[2] = {0.0f, 0.0f};
  __shared__ int sharedJ[threadsPerBlock];
  __shared__ float sharedMnmSource[2 * threadsPerBlock];
  __shared__ float sharedYnm[numCoefficients];
  __shared__ float sharedFactorial[2 * numExpansions];
  numInteraction = deviceOffset[blockIdx.x * offsetStride];
  for(i = 0; i < threadsPerBlock; i++){
    sharedJ[i] = 0;
  }
  for(j = 0; j < numExpansions; j++){
    for(k = 0; k <= j; k++){
      i = j * (j + 1) / 2 + k;
      sharedJ[i] = j;
    }
  }
  fact = 1.0;
  for(i = 0; i < 2 * numExpansions; i++) {
    sharedFactorial[i] = fact;
    fact = fact * (i + 1);
  }
  for(ij = 0; ij < numInteraction; ij++){
    jbase  = deviceOffset[blockIdx.x * offsetStride + 4 * ij + 1];
    dist.x = deviceOffset[blockIdx.x * offsetStride + 4 * ij + 2] * boxSize;
    dist.y = deviceOffset[blockIdx.x * offsetStride + 4 * ij + 3] * boxSize;
    dist.z = deviceOffset[blockIdx.x * offsetStride + 4 * ij + 4] * boxSize;
    for(i=0;i<2;i++) sharedMnmSource[2 * threadIdx.x + i] =
                     deviceMnmSource[2 * (jbase + threadIdx.x) + i];
    __syncthreads();
    cart2sph(rho, alpha, beta, dist.x, dist.y, dist.z);
    m2l_calculate_ynm(sharedYnm, rho, alpha, sharedFactorial);
    m2l_kernel_core(LnmTarget, sharedJ[threadIdx.x], beta,
                    sharedFactorial, sharedYnm, sharedMnmSource);
    __syncthreads();
  }
  ib = blockIdx.x * threadsPerBlock + threadIdx.x;
  for(i=0;i<2;i++) deviceLnmTarget[2 * ib + i] = LnmTarget[i];
}
\end{lstlisting}

\vspace{-4mm}

\section{Overview of Advanced Techniques}\label{advanced}
\vspace{\hup}

There are various techniques that can be used to enhance the performance of the treecode and \fmm. The \fmm\ presented in this article uses the standard translation operator for translating multipole/local expansions. As the order of expansion $p$ increases, the calculation increases as $\mathcal{O}(p^4)$ for this method. There are  alternatives that can bring the complexity down to $\mathcal{O}(p^3)$ \cite{ChengETal1999} or even $\mathcal{O}(p^2)$ \cite{GumerovDuraiswami2004}. In the code that we have released along with this article, we have included an implementation of the $\mathcal{O}(p^3)$ translation kernel by \cite{ChengETal1999} as an extension.  We have omitted the explanations in this text, however, and consider the advanced reader able to self-learn the techniques from the literature to understand the code.
Some other techniques that can improve the performance are the optimization of the order of expansion for each interaction \cite{Daschel2010}, the use of a more efficient \ML\ interaction stencil \cite{GumerovDuraiswami2008}, and the use of a treecode/\fmm\ hybrid, as suggested in \cite{ChengETal1999}. It is needless to mention that the parallelization of the code for multi-\gpu\ calculations \cite{HamadaNarumiYokotaYasuokaNitadoriTaiji09,Lashuketal2009} is an important extension to the treecode/\fmm\ on \gpu s.  Again, this is an advanced topic beyond the scope of this contribution.

When reporting the \gpu/\cpu\ speed up, it is bad form to compare the results against an un-optimized serial \cpu\ implementation.  Sadly, this is often done, which negatively affects the credibility of results in the field.  For this contribution, we have used a reasonable serial code in \textsc{C}, but it is certainly not as fast as it could be. For example, it is possible to achieve over an order of magnitude performance increase on the \cpu\ by doing single-precision calculations using \sse\ instructions with inline assembly code \cite{NitadoriYoshikawaMakino2010}. For those that are interested in the comparison between a highly tuned \cpu\ code and highly tuned \gpu\ code, we provide a highly tuned \cpu\ implementation of the treecode/\fmm\ in the code package that we release with this article.

\vspace{-4mm}

\section{Conclusions}
\vspace{\hup}

This contribution is a follow-on from the previous \emph{GPU Gems 3}, Chapter 31 \cite{NylandHarrisPrins2007}, where the acceleration of the all-pairs computation on \gpu s was presented for the case of the gravitational potential of $N$ masses.  We encourage the reader to consult that previous contribution, as it will complement the presentation we have given.
As can be seen in the results presented here, the cross-over point where fast $N$-body algorithms become advantageous over direct, all-pairs calculations is in the order of $10^{3}$ for the \cpu\ and in the order of $10^{4}$ for the \gpu.  Hence, utilizing the \gpu\ architecture moves the cross-over point upwards by one order of magnitude, but this size of problem is much smaller than many applications require.  If the application of interest involves, say, millions of interacting bodies, the advantage of fast algorithms is clear, in both \cpu\ and \gpu\ hardware.  With our basic kernels, about $15\times$ speedup is obtained from the fast algorithm on the \gpu\ for a million particles.  For $N=10^{7}$, the fast algorithms provide $150\times$ speedup over direct methods on the \gpu.  However, if the problem at hand requires small systems, smaller than $10^{4}$, say, one would be justified to settle for the all-pairs, direct calculation.

The main conclusion that we would like the reader to draw from this contribution is that constructing fast $N$-body algorithms on the \gpu\ is far from a formidable task.  Here, we have shown basic kernels that achieve substantial speedup over direct evaluation in less than 200 lines of \cuda code.  Expert-level implementations will, of course, be much more involved, and would achieve more performance. But a basic implementation like the one shown here is definitely worthwhile.

\medskip

We thank F.\ A.\ Cruz for various discussions that contributed to the quality of this article.

\vspace{\up}

\small
\bibliographystyle{plain}

\begin{thebibliography}{10}

\bibitem{Aarseth2003}
S.~Aarseth.
\newblock {\em Gravitational N-Body Simulations}.
\newblock Cambridge University Press, 2003.

\bibitem{Appel1985}
Andrew~W. Appel.
\newblock An efficient program for many-body simulation.
\newblock {\em SIAM J. Sci.\ Stat.\ Comput.}, 6(1):85--103, 1985.

\bibitem{BapatETal2009}
M.~S. Bapat, L.~Shen, and Y.~J. Liu.
\newblock Adaptive fast multipole boundary element method for three-dimensional
  half-space acoustic wave problems.
\newblock {\em Engineering Analysis with Boundary Elements},
  33(8--9):1113--1123, August--September 2009.

\bibitem{BarnesHut1986}
J.~Barnes and P.~Hut.
\newblock A hierarchical {$O(N \log N)$} force-calculation algorithm.
\newblock {\em Nature}, 324:446--449, December 1986.

\bibitem{Barnes1990}
J.~E. Barnes.
\newblock A modified tree code: Don't laugh; it runs.
\newblock {\em J.\ Comput. Phys.}, 87:161--170, 1990.

\bibitem{BoardSchulten2000}
J.~Board and K.~Schulten.
\newblock The fast multipole algorithm.
\newblock {\em Computing in Science and Engineering}, 2(1):76--79,
  January/February 2000.

\bibitem{BoardETal1992}
J.~A. Board, Jr., J.~W. Causey, J.~F. Leathrum, Jr., A.~Windemuth, and
  K.~Schulten.
\newblock Accelerated molecular dynamics simulation with the parallel fast
  multipole algorithm.
\newblock {\em Chem.\ Phys.\ Lett.}, 198(1--2):89--94, 1992.

\bibitem{ChengETal1999}
H.~Cheng, L.~Greengard, and V.~Rokhlin.
\newblock A fast adaptive multipole algorithm in three dimensions.
\newblock {\em J.\ Comput. Phys.}, 155:468--498, 1999.

\bibitem{DarveHave2004b}
E.~Darve and P.~Have.
\newblock Efficient fast multipole method for low-frequency scattering.
\newblock {\em J.\ Comput. Phys.}, 197:341--363, 2004.

\bibitem{Daschel2010}
H.~Daschel.
\newblock Corrected article: ``{An} error-controlled fast multipole method''.
\newblock {\em J. Chem.\ Phys.}, 132:119901, 2010.

\bibitem{DonepudiETal2003}
K.~C. Donepudi, J.-M. Jin, and W.~C. Chew.
\newblock A higher order multilevel fast multipole algorithm for scattering
  from mixed conducting/dielectric bodies.
\newblock {\em IEEE Transactions on Antennas and Propagation},
  51(10):2814--2821, 2003.

\bibitem{EthridgeGreengard2001}
F.~Ethridge and L.~Greengard.
\newblock A new fast-multipole accelerated {Poisson} solver in two dimensions.
\newblock {\em SIAM J. Sci. Comput.}, 23(3):741--760, 2001.

\bibitem{Fujiwara1998}
H.~Fujiwara.
\newblock The fast multipole method for integral equations of seismic
  scattering problems.
\newblock {\em Geophys.\ J. Intl.}, 133:773--782, 1998.

\bibitem{GorelickETal2006}
Lena Gorelick, Meirav Galun, Eitan Sharon, Ronen Basri, and Achi Brandt.
\newblock Shape representation and classification using the {Poisson} equation.
\newblock {\em IEEE Transactions on Pattern Analysis and Machine Intelligence},
  28:1991--2005, 2006.

\bibitem{GreengardLee1996}
L.~Greengard and J.-Y. Lee.
\newblock A direct adaptive {Poisson} solver of arbitrary order accuracy.
\newblock {\em J. Comp.\ Phys.}, 125:415--424, 1996.

\bibitem{GreengardRokhlin1987}
L.~Greengard and V.~Rokhlin.
\newblock A fast algorithm for particle simulations.
\newblock {\em J. Comput.\ Phys.}, 73(2):325--348, 1987.

\bibitem{GumerovDuraiswami2004}
N.~A. Gumerov and R.~Duraiswami.
\newblock {\em Fast multipole methods for the {H}elmholtz equation in three
  dimensions}.
\newblock Elsevier Series in Electromagnetism. Elsevier Ltd., 1st edition,
  2004.

\bibitem{GumerovDuraiswami2008}
N.~A. Gumerov and R.~Duraiswami.
\newblock Fast multipole methods on graphics processors.
\newblock {\em J. Comp.\ Phys.}, 227(18):8290--8313, 2008.

\bibitem{GumerovDuraiswami2009}
Nail~A. Gumerov and Ramani Duraiswami.
\newblock A broadband fast multipole accelerated boundary element method for
  the three dimensional {Helmholtz} equation.
\newblock {\em J. Acoust.\ Soc.\ Am.}, 125(1):191--205, 2009.

\bibitem{HamadaNarumiYokotaYasuokaNitadoriTaiji09}
T.~Hamada, T.~Narumi, R.~Yokota, K.~Yasuoka, K.~Nitadori, and M.~Taiji.
\newblock 42 {TFlops} hierarchical {N}-body simulations on {GPUs} with
  applications in both astrophysics and turbulence.
\newblock In {\em SC '09: Proceedings of the Conference on High Performance
  Computing Networking, Storage and Analysis}, pages 1--12, New York, NY, 2009.
  ACM.

\bibitem{KawaiETal1999}
A.~Kawai, T.~Fukushige, and J.~Makino.
\newblock \$7.0/{Mflops} astrophysical {$N$-body} simulation with treecode on
  {GRAPE-5}.
\newblock In {\em Supercomputing '99: Proceedings of the 1999 ACM/IEEE
  conference on Supercomputing}, New York, NY, USA, 1999. ACM.

\bibitem{Lashuketal2009}
I.~Lashuk, A.~Chandramowlishwaran, H.~Langston, T.~Nguyen, R.~Sampath,
  A.~Shringarpure, R.~Vuduc, L.~Ying, D.~Zorin, and G.~Biros.
\newblock A massively parallel adaptive fast-multipole method on heterogeneous
  architectures.
\newblock In {\em Proceedings of the Conference on High Performance Computing
  Networking, Storage and Analysis, SC '09}, pages 1--12, Portland, Oregon,
  November 2009.

\bibitem{LiuNishimura2006a}
Y.~J. Liu and N.~Nishimura.
\newblock The fast multipole boundary element method for potential problems: A
  tutorial.
\newblock {\em Engineering Analysis with Boundary Elements}, 30:371--381, 2006.

\bibitem{Liu2009}
Yijun Liu.
\newblock {\em Fast multipole boundary element method: Theory and applications
  in engineering}.
\newblock Cambridge University Press, 2009.

\bibitem{Makino1991}
J.~Makino.
\newblock Treecode with special-purpose processor.
\newblock {\em Publ.\ Astron.\ Soc.\ Japan}, 43:621--638, 1991.

\bibitem{Makino99}
J.~Makino.
\newblock Yet another fast multipole method without multipoles--pseudoparticle
  multipole method.
\newblock {\em J.\ Comput. Phys.}, 151:910--920, 1999.

\bibitem{MakinoTaiji1995}
J.~Makino and M.~Taiji.
\newblock Astrophysical {$N$-body} simulations on {GRAPE-4} special-purpose
  computer.
\newblock In {\em Supercomputing '95: Proceedings of the 1995 ACM/IEEE
  conference on Supercomputing}, page~63, New York, NY, USA, 1995. ACM.

\bibitem{MakinoTaiji1998}
Junichiro Makino and Makoto Taiji.
\newblock {\em Scientific Simulations with Special-Purpose Computers---the
  {GRAPE} Systems}.
\newblock John Wiley \& Sons Inc., 1998.

\bibitem{gems3}
Herbert Nguyen, editor.
\newblock {\em {GPU Gems 3}}.
\newblock Addison-Wesley Professional, 2007.
\newblock Available free online at
  \href{http://developer.nvidia.com/object/gpu-gems-3.html}{\textsf{http://dev%
eloper.nvidia.com/object/gpu-gems-3.html}}.

\bibitem{Nishimura2002}
N~Nishimura.
\newblock Fast multipole accelerated boundary integral equation methods.
\newblock {\em Appl.\ Mech.\ Rev.}, 55(4):299--324, 2002.

\bibitem{NitadoriYoshikawaMakino2010}
K.~Nitadori, K.~Yoshikawa, and J.~Makino.
\newblock Personal communication.

\bibitem{NylandHarrisPrins2007}
Lars Nyland, Mark Harris, and Jan Prins.
\newblock Fast {$N$}-body simulation with {CUDA}.
\newblock In {\em {GPU Gems 3}}, chapter~31, pages 677--695. Addison-Wesley
  Professional, 2007.

\bibitem{Rokhlin1990}
V.~Rokhlin.
\newblock Rapid solution of integral equations of scattering theory in two
  dimensions.
\newblock {\em J. Comp.\ Phys.}, 86(2):414--439, 1990.

\bibitem{SaguiDarden1999}
C.~Sagui and T.~A. Darden.
\newblock Molecular dynamics simulations of biomolecules: Long-range
  electrostatic effects.
\newblock {\em Ann.\ Rev.\ Biophys.\ Biomol.\ Struct.}, 28:155--179, 1999.

\bibitem{Taubes1997}
Gary Taubes.
\newblock The star machine.
\newblock {\em Discover}, 18(6):76--83, 1997.
\newblock available online at
  \href{http://discovermagazine.com/1997/jun/thestarmachine1148}{http://discov%
ermagazine.com/1997/jun/thestarmachine1148}.

\bibitem{TrefethenBau1997}
L.~N. Trefethen and D.~Bau, III.
\newblock {\em Numerical Linear Algebra}.
\newblock SIAM, Society for Industrial and Applied Mathematics, Philadelphia,
  PA, USA, 1997.

\end{thebibliography}

\end{document}